\documentclass[
twocolumn,
aps,nofootinbib,showpacs,showkeys,preprint
tightenlines,preprintnumbers,
,superscriptaddress
] {revtex4}

\usepackage{epsf,epsfig,subfigure,graphicx,amsmath,amssymb}
\usepackage{color}
\usepackage{float}

\newcommand{\dis}[1]{\begin{equation}\begin{split}#1\end{split}\end{equation}}

\newcommand\lsim{\mathrel{\rlap{\lower4pt\hbox{\hskip1pt$\sim$}}
    \raise1pt\hbox{$<$}}}
\newcommand\gsim{\mathrel{\rlap{\lower4pt\hbox{\hskip1pt$\sim$}}
    \raise1pt\hbox{$>$}}}

\newcommand\mplanck{M_{\rm P}}
\newcommand\ie{{\it i.e.}~}
\newcommand\fa{f_{a}}

\newcommand\gev{\,{\rm GeV}}

\newcommand\kev{\,{\rm keV}}

\newcommand\axino{{\tilde{a}}}

\newcommand\uonepq{U(1)$_{\rm PQ}$}

\newcommand{\Z}[1]{{\bf Z}}
\newcommand{\rhoz}{\rho_{\tiny\rm R}}
\newcommand{\az}{a_{\tiny\rm R}}
\newcommand\NDW{N_{\rm DW}}
\newcommand\NPQMSSM{N$_{\rm PQ}$MSSM}

\begin{document}

%%%%%%%%%%%%%%%%%%%%%%%%%%%%%%%%%%%%%%%%%%%%%%%%%%%%%%%%%%%%%%%%%
% End of Definitions and commands
%%%%%%%%%%%%%%%%%%%%%%%%%%%%%%%%%%%%%%%%%%%%%%%%%%%%%%%%%%%%%%%%%

\title{Mixing of axino and goldstino, and axino mass
}

\address{GIST College, Gwangju Institute of Science and Technology, Gwangju 500-712, Korea}
\author{Jihn E.  Kim$^{1}$, Min-Seok Seo}
\address{Department of Physics and Astronomy and Center for Theoretical
 Physics, Seoul National University, Seoul 151-747, Korea}

\begin{abstract}
Axino, related to the SUSY transformation of axion, can mix with goldstino in principle. This case is realized when some superfields carrying nonvanishing Peccei-Quinn charges develop both scalar VEVs and  F-terms. In this case, we present a proper definition of axion and axino. With this definition, we present the QCD axino mass in the most general framework, including non-minimal K\"ahler potential. The axino mass is known to have a hierarchical mass structure depending on accidental symmetries. With only one axino, if $G_A=0$ where $G=K+\ln|W|^2$, we obtain
$m_{\tilde a}=m_{3/2}$.
For $G_A\ne 0$, the axino mass depends on the details of the K\"ahler potential.
In the gauge mediation scenario, the gaugino mass is the dominant axino mass parameter. Therefore, we can take the theoretical QCD axino mass as a free parameter in the study of its cosmological effects, ranging from eV to multi-TeV scales, without a present knowledge on its ultraviolet completion.
\end{abstract}


\maketitle

%%%%%%%%%%%%%%%%%%%%%%%%%%%%%%%%%%%%%%%%
\section{Introduction}\label{sect:intro}
The lightest supersymmetric particle (LSP) with R-parity conservation
is absolutely stable and can contribute to the present energy
density of the universe as a dominant component of cold dark matter (CDM).  If the QCD axion solves the strong CP problem, its fermionic SUSY partner, {\em axino}, must be considered in the CDM estimate in cosmology. Firstly, it can be a natural candidate for CDM if it is the LSP~\cite{CKR00,CKKR01}. Second, if the axino is much heavier than the neutralino LSP, still it affects the mass density estimate of the neutralino LSP because the nonthermal decay of axino may dominate the estimate \cite{CKLS08}.
Therefore, it is of utmost importance to clarify what is the axino mass in a specific supergravity model.
The axino mass is obtained when SUSY is broken. A naive guess on the axino mass is of order $m_{3/2}$ since the axino mass is the soft term. But a leading loop corrections can reduce it to $\alpha m_{3/2}$. If this is suppressed, the next level hierarchical masses are arising from the gauge mediation and even some accidental symmetries can reduce it further to $m_{\rm S}(\fa/M_P), m_{\rm S}(\fa^2/M_P^2),\cdots $ \cite{ChunKN92}.
In fact, in the literature, axino mass has been considered in a vast mass regions from eV to trans-TeV \cite{Masiero84,RTW91,ChunKN92,CKR00,CKKR01, CKLS08,Steffen04,Huh09,ChoiCovi12,ChunLukas,NillesRaby82,Tamv82,Frere83,Strumia10,AxinoRevs}.

The axino mass depends on two symmetry breakings, the Peccei-Quinn (PQ) symmetry breaking and SUSY breaking. The effect of the PQ symmetry breaking must introduce a massless axion except from the contribution of the anomaly term and the effect of SUSY breaking introduces a mass parameter $m_{3/2}$. The contribution to the axino mass from the parameter $m_{3/2}$ is arising by the F-terms while the PQ symmetry breaking is given by the VEVs of scalar fields. In this paper, we study these two contributions in a most general form, and express the axino mass in terms of $m_{3/2}$ with the general K\"ahler form.

In global SUSY models, the axion $a$ is defined through the Peccei-Quinn(PQ) transformation for $\NDW=1$ fields \cite{KimRMP10},
\dis{
 \fa &e^{ia/\fa}=\sum_i v_i e^{i a_i/f_i}\\
& a\propto \sum_i v_i\Gamma_i \frac{a_i}{\fa}  \,,\label{eq:axiondef}
}
where $\Gamma_i,  v_i$ and $a_i$ are the eigenvalue of the PQ charge operator $\Gamma$, the vacuum expectation value(VEV) and the phase of $\phi_i$, respectively. The PQ direction of $a_i$ is $\Gamma_i\theta=\Gamma_i\frac{a}{\fa}$.

A prototype axion model with a global SUSY needs at least three chiral fields, to introduce a VEV breaking the PQ symmetry \cite{Kim83}. In this introduction, we do not introduce SUSY breaking, but only introduce the PQ symmetry and its breaking.
The superpotential having a global  PQ  symmetry is
\dis{
W=R(S_1S_2-f_a^2),\label{eq:PQZS1S2In}
}
where the PQ charges of $R, S_1$, and $S_2$ are $0,+1,$ and $-1$, respectively. Equation (\ref{eq:PQZS1S2In}) has an additional $R$-symmetry whose charges are 2, 0, and 0, respectively for  $R, S_1$, and $S_2$. To have the standard kinetic energy terms, the K\"ahler potential is taken as $K=RR^*+S_1S_1^*+S_2S^*_2$. Then, the potential $V$ is given by
\dis{
V=|S_1S_2-f_a^2|^2+|RS_2|^2+|RS_1|^2.
}
which is minimized at $\langle S_1S_2 \rangle=f_a^2$ and $\langle R\rangle=0$. To show the superTrace (STr) of $M^2$, we choose the fields near $\langle S_1 \rangle=\langle S_2 \rangle=f_a$,
\dis{&S_1=\frac{1}{\sqrt2}(\sqrt2f_a+\rho_1+ia_1)
\\
&S_2=\frac{1}{\sqrt2}(\sqrt2f_a+\rho_2+ia_2)
\\
&R=\frac{1}{\sqrt2}(\rhoz +i\az),}
where $\rho$'s are scalars and $a$'s are pseudoscalars.
The mass matrix for CP even scalars $(\rhoz, \rho_1, \rho_2)$ and CP odd scalars $(\az, a_1, a_2)$ are given by the same squared mass matrix,
\dis{f_a^2\left(\begin{array}{ccc}  2 & 0 &0 \\
      0 & 1 &1\\
      0&1&1
\end{array}\right)}
so eigenvalues are given by $(2f_a^2, 2f_a^2, 0)$. One CP odd scalar should be massless since it is the goldstone boson corresponding to the spontaneously broken PQ symmetry.

On the other hand, fermion mass matrix is given by
\dis{f_a\left(\begin{array}{ccc}  0 & 1 &1 \\
      1 & 0 &0\\
      1&0&0
\end{array}\right),\label{eq:axinoMassIn}
}
whose eigenvalues are $(\sqrt2 f_a, -\sqrt2 f_a, 0)$. Supersymmetry is not broken since $\langle V \rangle=0$, so should produce STr\,$M^2=0$,
\dis{
{\rm STr}\,M^2=m_{\rm CP even}^2+m_{\rm CP odd}^2-2m_{\rm fermion}^2=0.
}

In Eq. (\ref{eq:axinoMassIn}), the smaller mass is the axino mass, and one of the larger ones is the hypothetical goldstino mass. This larger one becomes exactly massless when the superHiggs mechanism is operative as will be discussed later.

The superpotential Eq. (\ref{eq:PQZS1S2In}) can be rewritten in terms of $R$,  the superfields $S=(S_1+S_2)/\sqrt2$, and the axion superfield $A=(S_1-S_2)/\sqrt2$ containing axion $(a_1-a_2)/\sqrt2$,

\dis{
W=R\left(\frac{S^2-A^2}{2} -f_a^2\right)\to -\frac12 R A^2\to 0.\label{eq:AandZIn}
}
The VEVs are $\langle R\rangle= \langle A\rangle= 0$ and $\langle S\rangle=\sqrt2 f_a$. After integrating out $R$, there does not exist low energy self interactions of $A$ due to the nonrenormalization theorem.

Here note that the zero mass eigenstates of scalar, pseudoscalar and fermion indicate that they are related by SUSY transformation and form the axion supermultiplet $A$. In general, the same mass eigenstates of scalar, pseudoscalar and fermion are not related by supersymmetry transformation and do not form a supermultiplet. For the supermultiplet condition, interactions must be supersymmetric also. In our case, however, the zero mass eigenstates form a supermultiplet $A$ which survives down to the low energy scale. This renders the nonlinear representations for $S_1=\varphi e^{A/f_a}$ and $S_2=\varphi e^{-A/f_a}$, which show explicitly the shift symmetry of the axion superfield.

%%%%%%%%%%%%%%%%%%%%%%%%%%%%%%%%%%%%%%%%%%%%%%%%%%%%%%%%%%%%%%
\section{The PQ symmetry in supergravity}\label{sec:AxinoTh}

The supersymmetrization of axion models
introduces a full axion supermultiplet $A$ which contains the
pseudoscalar axion $a$, its scalar partner {\em saxion} $s$, and their
fermionic partner axino $\axino$,
\dis{
  A=\frac{1}{\sqrt2}(s+ia)+\sqrt2 \axino \vartheta + F_A
  \vartheta\vartheta,\label{eq:axinodef}
}
where $F_A$ stands for an auxiliary field and $\vartheta$ for a Grassmann coordinate.

In the supersymmetric version of axionic models, the interactions of the saxion and the axino with matter are related by supersymmetry to those of the axion. Saxion and axino are better to accompany $a$ in Eq. (\ref{eq:axiondef}) to preserve SUSY. In Eq. (\ref{eq:axiondef}), $\fa$ is a VEV of some real scalar field which is called $\varphi_R$. Supersymmetrization of $\varphi_R$ needs its pseudoscalar partner $\varphi_I$, forming a complex scalar $\varphi$. When the PQ symmetry is not broken,  the PQ charged fields are of the $\varphi$ type, its phase changes when the PQ transformation is performed, and the fields are the real and the imaginary components of $\varphi$. On the other hand, if it is spontaneously broken, the PQ charge is not realized unitarilly but realized in the Nambu-Goldstone manner, \ie  the Goldstone boson $a$ is created, goes up to the phase as in Eq. (\ref{eq:axiondef}),  and the PQ symmetry is its shift symmetry. [The PQ symmetry is nonlinearly realized on the action to $a$.]   A nonzero SUSY breaking $F$-term of the  $\varphi$ type fields signals the PQ symmetry breaking also if it carry PQ charges. Therefore, the direction of axion does not necessarily coincide with the direction of axino which is going to be orthogonal to the goldstino which in turn is determined by the F-terms. In this paper, we define the axion and axino properly, and set up the formulae for the axino mass even in case that a mixing of axino with goldstino is present. Here, the axion component is still  defined by the coefficient of $\vartheta^0$ term since the F-term or the coefficient of $\vartheta^2$ term is auxilliary.

%%%%%%%%%%%%%%%%%%%%%%%%%%%%%%%%%%%%%%%%%%%%%%%%%%%%%%%%%%%%%%%%%%%
\subsection{Origin of axino-goldstino mixing}\label{subSec:originAxinoGoldMixing}

The PQ symmetry (as any global symmetries) in supersgravity has a meaning if both the superpotential $W$ and the K\"ahler potential $K$ respect the symmetry. Expansion of the K\"ahler potential in powers of $1/M_P$ leads to the following type,
\dis{
K= \sum_{I, J} f_I(\{\phi_i\}) g_J(\{\phi_j^*\})+{\rm h.c.}
}
where $\{\phi_i\}$ and $\{\phi_j^*\}$ are sets of holomorphic and anti-holomorphic fields, respectively. From the  K\"ahler potential, one can obtain its contribution to $V$ as
\dis{
V\in \int d^2 \vartheta  \int d^2 \bar{\vartheta} \sum_{I, J} f_I(\{\Phi_i\}) g_J(\{\overline{\Phi}_j\})+{\rm h.c.}
}
where $\{\Phi_i\}$ and $\{\overline{\Phi}_j\}$ are the superfields corresponding to $\{\phi_i\}$ and $\{\phi_j^*\}$, respectively.

Consider the leading term beyond the minimal term in $K$, for example $(1/M_P)H_u H_d X^*$. This must preserve the PQ symmetry so that $X$ carries the PQ charge $\Gamma(X)$ as the sum of $\Gamma(H_u)$ and  $\Gamma(H_d)$. Namely, the PQ symmetry is also broken by an $F$ term, \ie $X^*_F$ of $X^*$ \cite{GiuMas88}. This can be obtained from the superpotential as
\dis{
W\sim X_1X_2 X
}
where $\Gamma(X_1)+\Gamma(X_2)=-(\Gamma(H_u)+\Gamma(H_d))$. In this case, $\mu=-X_1X_2/M_P$ is obtained in Ref. \cite{KimNilles}.
This example shows that it is sufficient to consider the PQ charge carrying scalar components to pick up the axion component, and should not include the PQ charge carrying F-terms for a definition of the axion. Otherwise, we double count some components.

%%%%%%%%%%%%%%%%%%%%%%%%%%%%%%%%%%%%%%%%%%%%%%%%%%%%%%%%%%%%
\subsubsection{Introduction of $\varphi$ type fields}\label{subsubsec1}

In non-supersymmetric case, as axion is defined in Eq. (\ref{eq:axiondef}), and its property determined by the U(1)$_{\rm PQ}$ symmetry.  In the Wigner-Weyl (WW) realization of the PQ symmetry, the PQ charged fields transform as
\dis{
 \Phi_i\to e^{i\Gamma_i\theta}\Phi_i\\
\Gamma\Phi_i=\Gamma_i\Phi_i
}
where $\Gamma_i=-i\partial/\partial\theta$. In the Nambu-Goldstone (NG) phase, there appears a Goldstone boson $a$ which can be a combination of the original phase fields. Let $a$ be in $\Phi$ such that in the WW phase it is expanded as
\dis{
\Phi= \sum_i c_i X_i.\label{eq:PhiWW}
}
In the NG phase, the probability amplitude for $a$ to be in the phase of $X_i$ is $c_i=v_i/V_a$ where $V_a=\langle\Phi\rangle$ and  $v_i=\langle X_i\rangle$. The PQ operation on Eq. (\ref{eq:PhiWW})  is
\dis{
\Gamma_a\Phi=\sum_i c_i \Gamma_i X_i.\label{eq:PhiasLinearcom}
}

\begin{figure}[t]
  \begin{center}
  \begin{tabular}{c}
   \includegraphics[width=0.4\textwidth]{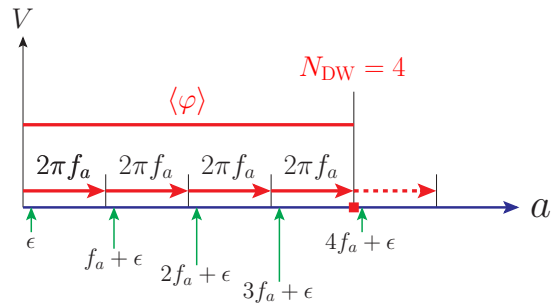}
  \end{tabular}
  \end{center}
  \caption{The definition of $\NDW$ in the Nambu-Goldstone phase. Here,  $\NDW=4$.
  }
\label{fig:DMW}
\end{figure}

When the PQ symmetry is realized in the NG manner by giving $X_i$ its VEV $v_i$, then the charge operator $\Gamma$ is not unitarilly realized, and then we must use the shift symmetry of the phase fields $a$, instead of the PQ symmetry operation $\Gamma$, with the original information on the eigenvalues $\Gamma_i$. So,  we apply the infinitesimal shift symmetry on $\sum_i c_i X_i$, $\delta a=\fa\delta\theta$ and $\delta a_i= \Gamma_i\fa\delta\theta$, and we use the relation proportional to $\delta\theta$ to code the spontaneous symmetry breaking.
For $\Phi\sim e^{i\Gamma_a a/\fa}$ where $\fa$ is determined by the axion-gluon-gluon anomaly term, then by acting the shift operation on $\Phi$ of Eq. (\ref{eq:PhiasLinearcom}) in the NG phase we obtain $\Gamma_a \Phi$ as $\sum_i(v_i/V_a)\Gamma_i X_i$ where $c_i $ is  $v_i/V_a$ which is the probability amplitude for $a$ to be in $X_i$. In  $\delta a_i= \Gamma_i\fa\delta\theta$, $\Gamma_i$ encodes the original degeneracy number of the phase of $X_i$ as the axion field $a$ completes one period in the NG phase. Representing $\Phi=(V_a+\rho_{\perp})e^{i\Gamma_a\, a/\fa}$
and  $X_i=(v_i+\rho_{i\perp})e^{i\Gamma_i a/\fa}$, where $\rho_\perp$ is in the perpendicular  direction to axion's Mexican hat valley and has mass proportional to $V_a$. Since the $\rho_{i\perp}$ masses are different, they should belong to different superfields. From this discussion, we obtain the  axion superfield $A$ in
\dis{
\Gamma_a\, \varphi_Ae^{  A/\fa}& \equiv  \sum_i  \frac{ v_i}{V_a}\Gamma_i\varphi_i\, e^{  A/ f_a} \label{eq:AxionSuperfield}
}
where $c_i=v_i/V_a$ is used. Note that we used $e^{  A/ f_a}$ on both sides in view of Fig. \ref{fig:DMW} since the axion shift of $2\pi\fa$ is fully accounted for by  $e^{  A/ f_a}$. Any integer cannot be multiplied or divided
in the exponent. Note that $\varphi_A$ is composed of two real fields $\rho_{\perp}$ and Im$\,\varphi_A$, so is  $\varphi_i$, and their VEVs are $\langle\varphi_A \rangle=V_a$ and $\langle\varphi_i\rangle=v_i$. Also, $A$ is composed of two real fields $s$ and $a$, and its VEV is vanishing  $\langle A\rangle=0$.
%%%%%%%%%%%%%%%%%%%%%%%%%%%%%%%%%%%%%%%%%%%%%%%%%%%%%%%%%%%%%%%%%%%%%%%%%%%%%%%%%%%
%%%%%%%%%%%%%%%%%%%%%%%%%%%%%%%%%%%%%%%%%%%%%%%%%%%%%%%%%%%%%%%%%%%%%

In the NG phase, we must state the domain wall number by the axion shift. In Fig. \ref{fig:DMW}, $\NDW=4$ is schematically shown. The decay constant $\fa$ is given by the coefficient of axion-gluon-gluon anomaly \cite{KimRMP10}. So, the axion potential returns to itself by a shift of $2\pi \fa$ \cite{KimPRP87}. The original domain wall number is given in Fig.  \ref{fig:DMW} by the length $\langle\varphi\rangle$.

Functions of the Nambu-Goldstone fields are multivalued. In the original field space, the field returns to itself after the $a$ shift of $2\pi \NDW\fa$.  If two axion directions have two domain wall numbers $n_1$ and $n_2$, the multiplicity of the vacua is the least common divisor
of $n_1$ and $n_2$ \cite{KimPRP87}. Thus, we can write $n_1=\NDW\beta_1$ and $n_2=\NDW\beta_2$, where $\beta_1$ and $\beta_2$ are relatively prime. So, Eq. (\ref{eq:AxionSuperfield}) can be written as
\dis{
\frac{1}{\NDW}\, \varphi_Ae^{  A/\fa}& \equiv  \sum_i  \frac{ v_i}{V_a}\frac{1}{\NDW\beta_i}\varphi_i\, e^{ A/ f_a} \label{eq:AxionDWN}
}
Therefore, we obtain
\dis{
a= \frac{\sum_i a_i/\beta_i}{\sqrt{\sum_i \beta_i^{-2}}}
}
whose literary form has the domain wall number 1 since  $\{\beta_i\}$ do not have a common divisor. From (\ref{eq:AxionDWN}), we obtain
\dis{
V_a^2=\sum_i \frac{v_i^2}{\beta_i}.\label{eq:Vasquare}
}

Now, consider the special case Eq. (\ref{eq:PQZS1S2In}) where in terms of three fields an ultraviolet completion is achieved, $W=R(S_1S_2-f_a^2)$ \cite{Kim83}. With the complete knowledge on $W$ of Eq. (\ref{eq:PQZS1S2In}),  $A$ and $\varphi$ are obtained in terms of $S_1$ and $S_2$ as $S_1=\varphi\, e^{A/f_a}$ and $S_2=\varphi\, e^{-A/f_a}$ \cite{Bae11}. Writing the low energy field $A$ in terms of the high energy fields $R, S_1$ and $S_2$ is not of much use in the region where only the low energy effective fields and the PQ quantum numbers of the original fields are known. In most cases, the information on $\varphi$ type fields is not needed.

%%%%%%%%%%%%%%%%%%%%%%%%%%%%%%%%%%%%%%%%%%%%%%%%%%%%%%%%%%%%
\subsubsection{Appearance of $A$ type fields in $W$}\label{subsubsec2}

For the K\"ahler potential, we can have any function of $A+A^*$.
But, a supersymmetric $W$ with $A$ is
\dis{
W(A)=0.
}
The U(1)$_{\rm PQ}$ invariance guarantees that the axion superfield $A$ does not appear in the superpotential.
But, below the SUSY breaking scale  we can introduce the soft terms in $W$ by introducing the auxilliary field $\Theta$ and respecting the shift symmetry,
\dis{
&W=M^3\Theta e^{\alpha A/f_a}\\
&\Theta=1+m_{\rm SUSY}\vartheta^2.
}
where $M$ is a parameter and $m_{\rm SUSY}=m_{\rm S}$ is the parameter describing  the SUSY breaking soft terms.  Then, $\ln|W|^2$ appearing in local asusy is invariant under the shift of $a$, with $A$  defined in Eq. (\ref{eq:AxionSuperfield}).

If there are more spontaneously broken global U(1)$_{A_i}(i=1,2,\cdots)$ symmetries, coaxions $a_i$ with the decay constant $f_i$ must respect the shift symmetries and we must consider the following $W$,
\dis{
W=M^3\Theta e^{\alpha A/f_a}\prod_{i=1,2,\cdots}e^{\alpha_i A_i/f_i} . \label{eq:WprodofA}
}

%%%%%%%%%%%%%%%%%%%%%%%%%%%%%%%%%%%%%%%%%%%%
\subsubsection{Comments on \NPQMSSM}

In the literature, models based on \NPQMSSM~ have been considered  \cite{KimNillSeo12,Yamaguchi11}.

In Ref. \cite{KimNillSeo12}, the original fields were used to show the existence of light field $X_{\rm ew}$ which must correspond to our $\varphi$ type field in the effective low energy theory framework. Also, by considering the original fields in a specific model, it was shown that $\varphi,\varphi^2$ and $\varphi^3$ are not present. In this effective theory, the PQ symmetry is broken, and the $\varphi$ type fields do not carry the PQ quantum number. So, $(\mu+\varphi)H_uH_d$ are the allowed interaction.  At the fundamental level, the $\mu$ term arises from the original fundamental fields, dictated by the PQ symmetry \cite{KimNilles}.
Counting the number of degrees of freedom, we double the fields $S_1$ and $S_2$ to
$\varphi_1 e^{A_1}$ and $\varphi_2 e^{A_2}$. As commented before, the light $A$ type field is $A=A_1-A_2$. $A_1+A_2$ becomes heavy. For the $\varphi$ type fields, only $X_{\rm ew}$ is light and the one orthogonal to it becomes heavy. So, at low energy, we have the exponential field $A$ and the $\varphi$ type field  $X_{\rm ew}$.
$X_{\rm ew}$ does not accompany a phase field. The coefficient of exponential of $A$ is very large, and $A$ does not appear explicitly in the superpotential. Axino and saxion are in $A$. So after integrating out our double counted fields, we end up with a superfield  $X_{\rm ew}$ and a superfield $A$. So $W$ of  Eq. (\ref{eq:PQZS1S2In}) has the same degrees at low energy in the NG phase. This is the way to write down the low energy theory corresponding to Eq. (\ref{eq:PQZS1S2In}). How  $X_{\rm ew}$ couples to the other light fields depends on the ultraviolet completion. Note also that the axion interaction depends on axion models \cite{KSVZ79,DFSZ81}.

On the other hand, Ref. \cite{Yamaguchi11} considered $\varphi,\varphi^2, \varphi^3$ and $\varphi H_uH_d$ terms without the $\mu$ term.

%%%%%%%%%%%%%%%%%%%%%%%%%%%%%%%%%%%%%%%%%%%%
\subsubsection{Comments on the model-independent axion}

The model-independent axion in superstring models is combined with the dilaton to make a supermultiplet \cite{Svrcek06},
\dis{
D= \frac{1}{g^2}+i\frac{a_{MI}}{8\pi M_P}\to  s +\frac{f_{MI}}{8\pi }e^{ia_{MI}/f_{MI}}
}
where $f_{MI}\sim 10^{16}\,\gev$ \cite{ChoiK85}, and $\langle s\rangle\simeq 2M_P$ is not the $\varphi$ type field. Because the corresponding U(1) is gauged,  $a_{MI}$ is absorbed to the U(1) gauge boson, and the U(1) symmetry remains as a global PQ symmetry below the scale $f_{MI}$. Only for this anomalous model-independent axion in string models, there is no accompanying $\varphi$ type field. Below $f_{MI}$, the resulting pseudo-Goldstone boson will accompany  a $\varphi$ type field.  We speculate that this model independent axion is the only place for the axion not accompanying its $\varphi$ type field.

%%%%%%%%%%%%%%%%%%%%%%%%%%%%%%%%%%%%%%%%%%%%%%%%%%%%%%%%%%%%%%%%%%%
\subsection{Goldstino, axion and axino}

The axion component is defined in Eq. (\ref{eq:axiondef}). So, whatever the non-vanishing PQ charge carrying F-terms are, the axion is properly defined only  by the PQ charge carrying $\vartheta^0$ terms. However, the nonvanishing F-terms define the goldstino component.

%%%%%%%%%%%%%%%%%%%%%%%%%%%%%%%%%%%%%%%%%%%%%%%%%%%%%%%%%%%%%%%%%%%%%%%%%%%%%%%%%%%
\begin{figure}[t]
  \begin{center}
  \begin{tabular}{c}
   \includegraphics[width=0.4\textwidth]{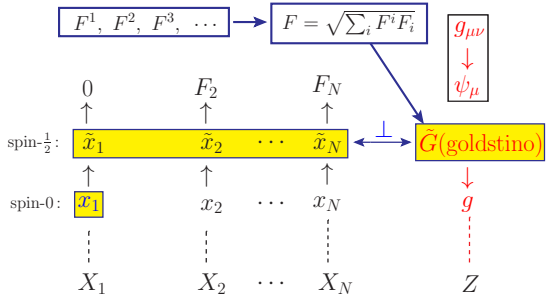}
  \end{tabular}
  \end{center}
  \caption{The case for more than one axino. It is ambiguous to choose the partner of the QCD axion. Normalization toward the canonical kinetic term is not depicted in the figure.
  }
\label{fig:ManyAxinos}
\end{figure}

Supersymmetry is spontaneously broken when the potential has nonzero VEV, $\langle V \rangle =\sum_i F^i F_i>0$ where $F^i \equiv K^{i{\bar j}}F_{\bar j}$.  Then, there should be a massless fermion, goldstino. In supergravity, it is absorbed to the longitudinal component of gravitino $\psi_\mu$ through the super-Higgs mechanism. The goldstino superfield, to which goldstino belongs, can be defined by
 \dis{
 Z=\sum_i \frac{F^i}{F} X_i, \label{eq:goldstDef}
 }
where $F=\sqrt{\sum_i F^iF_i}$which becomes the F-term of $Z$. Among $X_i$, the axion superfield is defined by the PQ charges of $X_i$. All the other chiral fields orthogonal to $A$ are called {\it coaxino} directions.
Then, we can consider two cases in which the axion superfield $A$ allows
  \begin{itemize}
  \item{ $F_{A} \ne 0$, or}
  \item{$F_{A} =0$, but $F^A\ne 0$ from K\"ahler mixing with other SUSY breaking fields.}
  \end{itemize}
This case is shown in Fig. \ref{fig:ManyAxinos}.

With the canonical K\"ahler potential, $F^i$ is just $F_i^*$, and the goldstino fermion gives the SUSY breaking direction exactly. On the other hand, in the presence of a K\"ahler mixing, $F^i$ is nonzero even though $F_i=0$ and seems to contribute to the goldstino. However, the physical goldstino, defined by the zero mass eigenstate with the canonical kinetic term, does not contain such a state. This can be shown as follows. In supergravity, fermion kinetic and mass terms are given by
\dis{
e^{-1}{\cal L}=-iG_{i{\bar j}}\bar{\psi'}^{\bar j} \bar{\sigma}_\mu {\cal D}^{\mu} {\psi'}^i +\frac12 m'_{ij}{\psi'}^i{\psi'}^j+\frac12 {m'}^\dagger_{{\bar i}\,{\bar j}}{\bar{\psi'}}^{\bar i}{\bar{\psi'}}^{\bar j} \label{eq:massterm}
}
where ${\cal D}_\mu$ is a general covariant derivatives and $m'_{ij}=m_{3/2}[\nabla_iG_j+(1/3M_P^2)G_iG_j]$ is the fermion mass with the goldstino field moded out. The primed fields and the primed mass matrix are in the interaction basis. To obtain the physical states, we first make the kinetic terms canonical and then diagonalize the mass matrix. With the redefinition of
${\psi'}^i=V^i\,_{a}\psi^a$, Eq. (\ref{eq:massterm}) is  written as
\dis{
e^{-1}{\cal L}&=-i[{V^\dagger}_{\bar a}\,^{\bar j}G_{i{\bar j}}V^i\,_{b}]\bar{\psi}^{\bar a} \bar{\sigma}_\mu {\cal D}^{\mu} {\psi}^b \\
 &+\frac12 [V^Tm'V]_{ab}{\psi}^a{\psi}^b+\frac12 [V^\dagger{m'}^\dagger V^*]_{{\bar a}{\bar b}}{\bar{\psi}}^{\bar a}{\bar{\psi}}^{\bar b}+\cdots
 \\
 &=-i\bar{\psi}^{a} \bar{\sigma}_\mu {\cal D}^{\mu} {\psi}^a +\frac12 m_{ab}{\psi}^a{\psi}^b+\frac12 m^\dagger_{{\bar a}{\bar b}}{\bar{\psi}}^{\bar a} {\bar{\psi}}^{\bar b}+\cdots.
 }
Requiring the canonical kinetic terms as ${V^\dagger}_{\bar a}\,^{\bar j}G_{i{\bar j}}V^i\,_{b}=\delta_{{\bar a}b}$, or ${V^T}_{b}\,^iG_{ i {\bar j}} {V^*}^{\bar j}\,_{\bar a}=\delta_{b{\bar a}}$, the mass in this basis is $m_{ab}=[V^Tm'V]_{ab}$ which we have to diagonalize.

The matrix $V$ can be written in the form of $V=US$ where $U$ is a unitary matrix and $S$ is a scaling matrix. Taking the inverse of the canonical kinetic term condition, ${V^T}_{b}\,^iG_{i\bar{j}}{V^*}^{\bar j}\,_{\bar a}=\delta_{b{\bar a}}$, we obtain
\dis{
&(V^{*-1})^{\bar a}\,_{\bar j}G^{{\bar j} i}(V^{T-1})_{i}\,^{b}
\\
&=(S^{-1}U^T)^{\bar a}\,_{\bar j} G^{\bar{j}i}(U^*S^{-1})_i\,^b =\delta^{{\bar a}b}.
}
Taking the complex conjugation, we obtain
 \dis{(S^{-1}U^\dagger)^a\,_{j}G^{j\bar{i}}(US^{-1})_{\bar i}\,^{\bar b} =\delta^{a{\bar b}},}
 or equivalently,
 \dis{(S^{-1}U^\dagger)^a\,_{j}G^{j\bar{i}}=(SU^\dagger)^{a\bar i} \label{eq:redcond}}

On the other hand, $m'_{ij}$ has zero eigenvalue in the direction of $G^i=G^{i{\bar j}}G_{\bar j}$. Since $m'=(V^T)^{-1}mV^{-1}$$=(U^*S^{-1})m(S^{-1}U^\dagger)$, $m'_{ij}G^j=0$ implies  $m_{ab}(S^{-1}U^\dagger)^b\,_{j}G^j=0$. So, $(S^{-1}U^\dagger)^b\,_{j}G^{j{\bar k}}G_{\bar k}$ is the goldstino direction. Plugging Eq. (\ref{eq:redcond}) into this, we obtain
\dis{
(S^{-1}U^\dagger)^a\,_{j}G^{j{\bar k}}G_{\bar k}=(SU^\dagger)^{a \bar j}G_{\bar j}.
}

We have rotated fermion as ${\psi'}^i=V^i\,_{a}\psi^a$ to make the kinetic term canonical. So, in the new (physical) basis, $\psi^a=(V^{-1})^a\,_{i}(\psi')^i=(S^{-1}U^\dagger)^a\,_{i}(\psi')^i$. Note that $S$ is a real scaling diagonal matrix. Rotation of the direction is determined by $U$. SUSY breaking direction $G_i$  defined in the interation basis $\psi'$ is rotated by $U$ in the new basis $\psi$. Then ${U^\dagger}^{a\bar i}G_{\bar i}$ is just supersymmetry breaking direction of $G_{\bar i}$, not $G^i$.

As indicated within the yellow box in Fig. \ref{fig:ManyAxinos}, there is an ambiguity in identifying the mass eigenstate axino corresponding to the QCD axion.

\begin{figure}[t]
  \begin{center}
  \begin{tabular}{c}
   \includegraphics[width=0.4\textwidth]{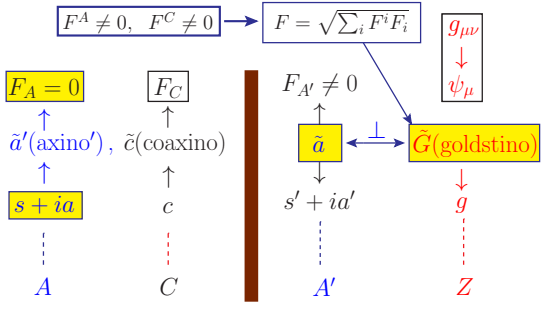}
  \end{tabular}
  \end{center}
\caption{The case  $F_A=0$ for the axion (blue) and goldstino (red) multiplets. The axion direction $a$ is defined by the PQ symmetry through Eq. (\ref{eq:axinodef}) and the goldstino ($\tilde G$) and axino ($\tilde a$) directions are defined by the fermion mass eigenvalues.
Normalization toward the canonical kinetic term is not depicted in the figure.}
\label{fig:FAzero}
\end{figure}

\begin{figure}[t]
  \begin{center}
  \begin{tabular}{c}
   \includegraphics[width=0.4\textwidth]{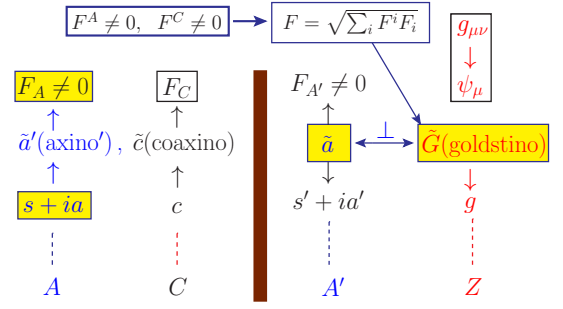}
  \end{tabular}
  \end{center}
  \caption{The same as Fig. \ref{fig:FAzero} except for $F_A\ne 0$.
  }
\label{fig:FAnonzero}
\end{figure}

When there exists only one axino, there is no ambiguity in identifying the mass eigenstate QCD axino, because it must be the orthogonal state to the goldstino. For two fermions, we can consider two cases separately as shown in Fig. \ref{fig:FAzero} for $F_A=0$ and Fig. \ref{fig:FAnonzero} for $F_A\ne 0$. Even for one axino and goldstino case, the one beyond the axion multiplet is called the coaxino as indicated in the left-hand side of
Figs. \ref{fig:FAzero} and \ref{fig:FAnonzero} with a thick brown bar.  These relatively simple cases will be discussed explicitly in Sec. \ref{sec:Calculation} where we calculate the axino mass by the gravity mediation.
In the regrouping of Figs. \ref{fig:FAzero} and \ref{fig:FAnonzero}, we violated the supermultiplet condition, $QA=\tilde a$, and may make  an error in estimating the axino mass by an order  O($m_{3/2}^2/M_P$).

%%%%%%%%%%%%%%%%%%%%%%%%%%%%%%%%%%%%%%%%%%%%%%%%%%%%%%%%%%%%%%%%%%%%%%%%%%%%%%%%%%%
%%%%%%%%%%%%%%%%%%%%%%%%%%%%%%%%%%%%%%%%%%%%%%%%%%%%%%%%%%%%%%%%%%%%%%%%%

The goldstino multiplet $Z$ is defined by the total $F^*$ term,
\dis{
Z=\frac{F^A}{F}A+\sum_{i \ne A} \frac{F^i}{F}X_i  \equiv \frac{F^A}{F}A+\frac{F^C}{F}C,
}
where $C$ is the sum of SUSY breaking coaxino, orthogonal to the axino superfield. Their bosonic and fermionic components  are not the components of the original superfields. Since axion and goldstino are finally defined after SUSY is broken, the mismatch is generic.  In view of Figs. \ref{fig:FAzero} and \ref{fig:FAnonzero}  introducing one coaxino, we can regroup the axion and coaxino multiplets as
\dis{
&A=(A, \tilde{a}', F_A)\\
 &C=(c', \tilde{c}, F_C)
}
where the scalar component of the axion multiplet is $A=s'+ia$. $a'$ in Figs. \ref{fig:FAzero} (and also in \ref{fig:FAnonzero})
is not a mass eigenstate since the identification of $\tilde a\perp \tilde G$ does not care about its superpartner $a'$.
The scalar potential is given by
\dis{
V=M_P^2e^{G/M_P^2}[G^{i{\bar j}}G_iG_{\bar j}-3M_P^2],
}
and pseudoscalar $a$ in $A$ does not appear in $V$ since $G$ is a function of $(1/2)(A+A^*)=s'$. It is a massless eigenstate, and therefore the pseudoscalar mass matrix is already diagonalized in the left-hand side  of the brown bar.   On the other hand, the fermion mass matrix should be diagonalized in the right-hand side  of the brown bar. One eigenvalue in the direction of $F^i$, $(F^A, F^C)$
is massless along the direction of $Z$ which is goldstino.  After making the fermion kinetic term canonical, goldstino indicates the direction $(F_A, F_C)$. The remaining eigenvalue is interpreted as the axino mass.
If $F_A \ll F_C$, $ \tilde{a}\simeq \tilde{a}'-(F^A/F)\tilde{Z} $ and the mass eigenstate is axino-like.

   In general, however, after the SUSY breaking, the mass eigenstates are
\dis{{\rm Scalar}:&~~~s,~{\rm Re}\,c \\
 {\rm Pseudoscalar}:&~~~a,~{\rm Im}\,c={\rm Im}\,c' \\
  {\rm Fermion}:&~~~\tilde{a}=\tilde{a}'-(F^A/F)\tilde{Z} ,~{\tilde Z},
}
where $s$ and $c$ are the mass eigenstates after diagonalizing the mass matrix in the $(s', c')$ basis. In the last step, the shift symmetry of $a$ must be invoked to guarantee the massless axion.

%%%%%%%%%%%%%%%%%%%%%%%%%%%%%%%%%%%%%%%%%%%%%%%%%%%%%%%%%%%%%%%%%%%%%%%%%%%%%%%%%%%

This process makes sense when we consider the axion  interaction with gluino
\dis{
\frac{1}{f_a}\int d^2 \vartheta A\, {\cal W}^\alpha {\cal W}_\alpha\,. \label{eq:AxionGluon}
}
If $F_A =0$ as in Fig. \ref{fig:FAzero}, the gluino mass does not get a contribution from the axion multiplet. However, if $F_{A}\ne 0$,  the gaugino mass should be studied carefully in a specific model. This is true also if the K\"ahler potential has a large mixing term.

In our language, after the goldstino $\tilde G$ is absorbed to $\psi_\mu$, we can write the $\psi_\mu$ interaction as\cite{Deser77},
\dis{
\frac{m_{3/2}}{2}\,\psi_\mu \sigma^{\mu\nu}\psi_\nu=\int d^2\vartheta \frac{G}{2 \sqrt3 M_P}\,\psi_\mu \sigma^{\mu\nu}\psi_\nu.
}
So, before hiding the goldstino into $\psi_\mu$, any chiral field interaction in supergravity has a coupling suppressed by $1/M_P$,
\dis{
\frac{{\rm (coupling)\,}\Phi_i'}{M_P}.
}
Therefore, the coupling of $A'$ is suppressed by $M_P$, not by $f_a$, in Eq. (\ref{eq:AxionGluon}).
After absorbing the goldstino into $\psi_\mu$, any other chiral fields are orthogonalized not to have an F-term but its supergravity coupling is suppressed by $1/M_P$ again, and the goldstino superfield $G$ couples to a gauge supermultiplet as
\dis{
\frac{{\rm (coupling)}}{M_P}\int d^2 \vartheta\, Z\, {\cal W}^\alpha {\cal W}_\alpha
}
where $F_G=F$ gives the gaugino mass. In addition, the axino interaction with gluino is
\dis{
 &\frac{1}{f_a}\int d^2\vartheta A\,{\cal W}^\alpha {\cal W}_\alpha\\
  &\to \frac{1}{f_a}  (\tilde a-\epsilon_a \tilde G)\,  ({\rm gluino})^a({\rm gluon})^a\,\\
  &\to \frac{1}{f_a}\tilde a  ({\rm gluino})^a({\rm gluon})^a\,\\
  &\hskip 0.5cm -\epsilon_a  \frac{1}{f_aM_P}(\partial^\mu\psi_\mu) ({\rm gluino})^a({\rm gluon})^a\,.
}
So axino interaction implies some accompanying gravitino interaction,
suppressed by the product $f_a M_P$. This makes sense since it must respect the spontaneous symmetry breaking suppressed by $\fa$ and the super-Higgs mechanism suppressed by $M_P$.

%%%%%%%%%%%%%%%%%%%%%%%%%%%%%%%%%%%%%%%%%%%%%%%%%%%%%%%%%%%%%%%%%%%%%
\subsection{Axino mass with SUSY diagrams}\label{subsec:Loops}
%%%%%%%%%%%%%%%%%%%%%%%%%%%%%%%%%%%%%%%%%%%%%%%%%%%%

\begin{figure}[t]
  \begin{center}
  \begin{tabular}{c}
   \includegraphics[width=0.4\textwidth]{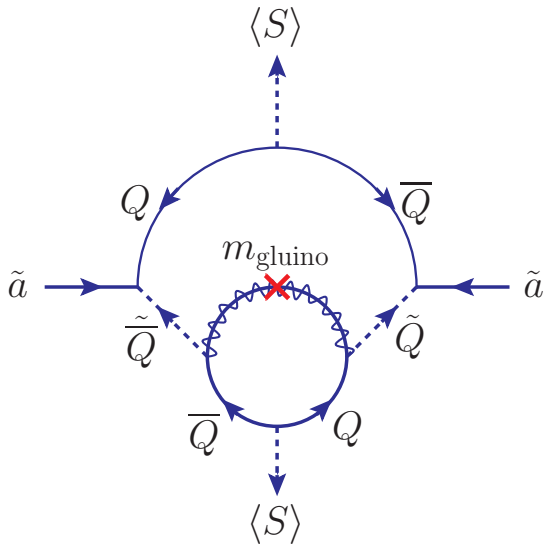}
  \end{tabular}
  \end{center}
  \caption{The two-loop axino mass in the KSVZ model.
  }
\label{fig:AxinoKSVZ}
\end{figure}
%%%%%%%%%%%%%%%%%%%%%%%%%%%%%%%%%%%%%%%%%%%%%%%%%%%%%%%%%%%%%%%%%%%

In Eq. (\ref{eq:axinoMassIn}), the smaller mass was shown as the axino mass, and one of the larger ones is the hypothetical goldstino mass. This larger one becomes exactly massless when the superHiggs mechanism is operative as discussed in the previous section. This phenomenon can be compared to the positive complex scalar mass splitting into one Higgs boson mass plus a goldstone boson mass when the Higgs mechanism is in operation. The correspondence is what  the axino mass to the goldstino is what the Higgs boson mass to the goldstone boson mass.

But there is also a non-gravitational contribution to the axino mass. Even if the axion is massless, both fermion partners are massive as shown in Eq. (\ref{eq:axinoMassIn}). This is a supersymmetric mass where fermionic masses are split differently from those of bosons. As an example, consider a KSVZ coupling
\dis{
W=-f_Q S\, \overline{Q}\, Q.
}
There exists a two-loop diagram generating the axino mass as shown in Fig. \ref{fig:AxinoKSVZ}. The mass estimate is
\dis{
m_{\tilde a}\sim \frac{f_Q^2 g_c^2}{(4\pi^2)^2}m_{\rm gluino}\simeq\,
\frac{\alpha_{f_Q}\alpha_c}{\pi^2}m_{\rm gluino}.
}
So, the axino mass contains the contribution
\dis{
m_{\tilde a}=  \sum_{a= {\rm gaugino}} \xi_{a\,} m_{1/2\,,a }+\cdots
}
where  $m_{1/2\,,a }$ are the gaugino masses. These contributions give masses of order up to probably 10  GeV.

In the gravity mediation scenario, there exists another parameter $m_{3/2}$ which can be much larger than 10 GeV. Without the axino-goldstino mixing in the K\"ahler potential, the axino direction is the same as that of axion and the superpotential determines the axino mass. It means that axino mass arises from loop diagrams as in the Fig. \ref{fig:AxinoKSVZ} example. Therefore, without the axino-goldstino mixing in the K\"ahler potential, axino mass is not going to be larger than 10 GeV. Thus, a very heavy axino mass is possible only if there is a significant $A-Z$ mixing in the K\"ahler potential.

If we consider the anomaly mediation, there would be additional contributions from breaking the conformal symmetry which appear as anomalous dimensions in the superpotential terms \cite{AnomMed02}. Generally, they are proportional to $\gamma_{{\rm anom}\, I}\,m_{3/2}$ for the term $W_I$ in the superpotential. Thus, the axino mass has contributions from all the above cases and expressed as
 \dis{
m_{\tilde a}\,= & \left(\xi_{\rm goldstino}+\sum_{I= {\rm terms~in~}W} \xi^{\rm anom}_{I}\right)m_{3/2}\\
& + \sum_{a= {\rm gaugino}} \xi_{a\,} m_{1/2\,,a } .\label{eq:axinomassxis}
}
We expect that $\xi^{\rm anom}_{I}$ is of order O($10^{-2}$). An example has been discussed in \cite{Yamaguchi02}.

%%%%%%%%%%%%%%%%%%%%%%%%%%%%%%%%%%%%%%%%%%%%%%%%%%%%%%%%%%%%%%
\subsection{With accidental symmetries}

In \cite{ChunKN92}, the possiblility of keV axino was discussed in case the superpotential has an accidental symmetry. The keV \cite{RTW91} and even eV \cite{Masiero84} range axino masses are possible with some accidental symmetries. The accidental symmetries may forbid the leading order masses of the scales $m_{3/2}$ and $m_{1/2\,,a }$.

In the gauge mediation scenario, $m_{3/2}$ is negligible and the axino-goldstino mixing does not give a  significant contribution. Then, the loops may give the dominant contribution. But the accidental symmetry may forbid diagrams of the form of Fig. \ref{fig:AxinoKSVZ}. The superpotential may introduce a nonrenormalizable term suppressed by $M_P$, and the expansion parameter is $f_a/M_P\sim 10^{-7}$. Thus, the axino mass of Fig. \ref{fig:AxinoKSVZ} is further suppressed by $\sim 10^{-7}$ and we expect $10\,\gev\cdot 10^{-7}\simeq 1\,\kev$. If it is further suppressed, then the estimated axino mass is of order $10^{-3}\,$eV.

In the gravity mediation scenario, $m_{3/2}$ is a TeV scale and the axino mass depends on the K\"ahler potential. Without the axino-goldstino mixing, it was commented in Subsec. \ref{subsec:Loops}, and  the discussion with some accidental symmetries is the same as the above paragraph.

%%%%%%%%%%%%%%%%%%%%%%%%%%%%%%%%%%%%%%%%%%%%%%%%%%%%%%%%%%%%%%%%%%%%%
\section{Parametrization with nonminimal K\"ahler form}

Chun and Lukas studied the axino mass with the minimal K\"ahler form  \cite{ChunLukas}. Here we go beyond the minimal K\"ahler form, work with the PQ symmetry realized in the NG manner, and include the effects of $F$ terms of the PQ charged fields which affect the axino component.

Equation (\ref{eq:AxionSuperfield}) is our definition of axion superfield, and
the K\"ahler potential must respect the shift symmetry of axion. Therefore, the lowest order terms in the K\"ahler potential with some mixing with SUSY breaking coaxino $C$ are\footnote{The $M$ term contributes to the saxion derivative in the Lagrangian as $M\partial^2 s$, which has no effect to the axino mass.}
\dis{
K=&\frac12(A+A^*)^2+\epsilon (A+A^*)(C+C^*) \\
 &+CC^*+M(A+A^*).\label{eq:MixKaehler}
 }

The SUSY breaking can be introduced in terms of Kim's generalized form \cite{Kim83} of the Polonyi one \cite{Polonyi77}. Here, however, we will parametrize the SUSY breaking just by an auxilliary holomorphic constant $\Theta$,
\dis{
\Theta=1+ m_{\rm S}\vartheta^2.\label{eq:Theta}
}
In view of the discussion of Subsubsec. \ref{subsubsec2}, if there are coaxions then
the superpotential can be taken as
\dis{
W(C)=\frac{C^4}{M_P}\Theta+\cdots
}
with $\langle W(C)\rangle=M^3\sim (10^{13}\,\gev)^3$.

%%%%%%%%%%%%%%%%%%%%%%%%%%%%%%%%%%%%%%%%%%%%%%%%%%%%%%%%%%%%%%
\subsection{Local SUSY with one axino}\label{sec:Calculation}

The most important requirement is that goldstino is defined in the vanishing cosmological constant(CC) vacuum, satisfying the U(1) invariance condition. Thus, in calculating the axino mass, we satisfy the following three conditions:
\begin{itemize}
\item[({\it i})] The vanisihing CC condition,
\dis{
\hskip -1cm G^{i{\bar j}}G_iG_{\bar j}=3M_P^2,\label{eq:Cond1}
}
where where $G=K+M_P^2 \ln|W|^2$.
\item[({\it ii})]  The vacuum stabilization condition,
\dis{
\hskip -1cm G^{j {\bar k}}G_{\bar k}\nabla_i G_j+G_i=0.\label{eq:Cond2}
}

\item[({\it iii})]  For the U(1) invariance condition, we use
\dis{&K=K(A+A^*, C,C^*)  \\
 &W=\Theta \, e^{\alpha A/f_a}W(C)\label{eq:WofA}
. }
If there are more than one coaxino, we have
\dis{
W=W(C)\, e^{\alpha A/f_a}\times  e^{\alpha A_1/f_1}\times\cdots
}
The superpotential in (\ref{eq:WofA}) preserves the shift symmetry of $A$ since
in  $G=K+\ln|W|^2$, the $|W|^2$ part is read as
$|W|^2=|W(C)|^2\, \Theta\, e^{\alpha(A+A^*)/f_a}.$

\end{itemize}
Now, let us calculate the axino mass given by
\dis{
     m=m_{3/2}[\nabla_iG_j+\frac13G_iG_j]
     }
for two classes of $\langle C\rangle=0$ and $\langle C\rangle\ne 0$.

%%%%%%%%%%%%%%%%%%%%%%%%%%%%%%%%%%%%%%%%%%%%%%%%%%%%%%%%%%%%%%%%%%%%%%%%%%%%%%%%
\subsubsection{Case for $G_A=0$ and $G^A\ne 0$}

As an example for $G_A=0$ but for $G^A\ne 0$, we consider
\dis{
&K=\frac12(A+A^*)^2+CC^*+\epsilon(A+A^*)(C+C^*)\\
&\quad \quad W=e^{\alpha A/f_a}W(C)
}
which is an example for $G_A=0$ but $G^A\ne0$; so goldstino has some axino component.
From $G=K+M_P^2\ln|W/M_P^3|^2$,   we obtain
\dis{
&G_A=(A+A^*)+\epsilon(C+C^*)+\frac{\alpha}{f_a}
\\
&G_C=C^*+\epsilon(A+A^*)+M_P^2\frac{W_C}{W}
}
from which the K\"ahler metric elements and their inverse elements are given by
\dis{
&G_{A\bar{A}}=1,~G_{A\bar{C}}=\epsilon,~G_{C\bar{A}}=\epsilon,~G_{C{\bar C}}=1,\\
&G^{A\bar{A}}=\frac{1}{1-\epsilon^2},~G^{A\bar{C}}=-\frac{\epsilon}{1-\epsilon^2},\\
&G^{C\bar{A}}=-\frac{\epsilon}{1-\epsilon^2},~G^{C{\bar C}}=\frac{1}{1-\epsilon^2}.\label{eq:metrics}
}
The $G_{ij}$ elements which are needed for the mass matrix elements are given by
\dis{
&G_{AA}=1,~~G_{AC}=\epsilon,\\
&G_{CC}=M_P^2\left[\frac{W_{CC}}{W}-\Big( \frac{W_C}{W}\Big)^2\right].
}
Since $G_{i\bar{j}}$ are constants, $G_{ij{\bar k}}=0$. For this reason, the Christoffel symbol $\Gamma^i_{jk}\equiv G^{i\bar{l}}G_{jk{\bar l}}=0$, and hence the K\"ahler geometry is flat.

The vacuum conditions ({\it i}) and ({\it ii}) are given by
    \dis{
    &G^{i{\bar j}}G_iG_{\bar j}=3M_P^2,\\
    & G^{j{\bar k}}G_{\bar k}\nabla_iG_j+G_i=0.
    }
Since we defined the axion superfield $A$ as the supersymmetrization of Goldstone boson $a$, the VEV $\langle A\rangle$ is zero as discussed in Sec. \ref{sec:AxinoTh}.
Now, we calculate for two cases of $\langle C\rangle=0$ and $\langle C\rangle\ne 0$.

%%%%%%%%%%%%%%%%%%%%%%%%%%%%%%%%%%%%%%
\vskip 0.3cm
\noindent { (i) $\langle C\rangle=0\,$}:
\vskip 0.1cm
Three vacuum conditions of Eqs. (\ref{eq:Cond1}) and (\ref{eq:Cond2}) give the following,
\dis{
   &2\frac{\alpha}{f_a}=0,\\
   &\frac{1}{1-\epsilon^2}\Big|\frac{W_C}{W}\Big|^2=\frac{3}{M_P^2}\\
&\frac{1}{1-\epsilon^2}\Big(\frac{W_{CC}}{W}M_P^2-\epsilon^2\Big)=
2\frac{ W_C/ W}{ W^*_{\bar C}/W^*}.
}
The first of these requires $\alpha=0$, and $W$ does not depend on $A$. Thus, in calculating $G_A=0$ the $\ln|W|^2$ part of $G$ does not contribute and we cannot constrain the superpotential from the $G_A=0$ condition.

 The axino mass can be obtained from the mass matrix in the $(A, C)$ basis as,
\dis{
m&=m_{3/2}\left(\begin{array}{cc}  1  &\epsilon \\
      \epsilon & M_P^2\Big[\frac{W_{CC}}{W}-\frac23\Big(\frac{W_C}{W}\Big)^2 \Big]\\
\end{array}\right)\\
&=m_{3/2}\left(\begin{array}{cc}  1  &\epsilon \\
      \epsilon & \epsilon^2\\
\end{array}\right)}
 The vacuum conditions determine the (22) element to be $\epsilon^2$.

To make fermion mass term canonical, we have to redefine fermion as $\psi \to V\psi$ where
\dis{V=US=\frac{1}{\sqrt2}\left(\begin{array}{cc}  1& -1  \\
     1 &1
\end{array}\right)\cdot
\left(\begin{array}{cc}  \frac{1}{\sqrt{1+\epsilon}}&0  \\
     0&\frac{1}{\sqrt{1-\epsilon}}
\end{array}\right) \label{eq:Vexample}}
and the mass matrix $m=V^T(\nabla_i G_j+(1/3M_P^2)G_iG_j)V$ is given by
\dis{\frac{m}{m_{3/2}}=\frac12\left(\begin{array}{cc}  1+\epsilon& -\sqrt{1-\epsilon^2}  \\
     -\sqrt{1-\epsilon^2} &1-\epsilon
\end{array}\right).\label{eq:60}
}
Note that $|\epsilon| < 1$ is requred to make fermion kinetic term positive definite.
The eigenvalues of Eq. (\ref{eq:60}) are 0 and $m_{3/2}$. This confirms that the vanishing goldstino mass comes out right.  The coefficient of $m_{3/2}$ in the axino mass, $\xi_{\rm goldstino}$, is 1. The axino mass independence of $\epsilon$ shows that goldstino is defined in the direction of $G_C$ only.

%%%%%%%%%%%%%%%%%%%%%%%%%%%%%%%%%%%%%%
\vskip 0.3cm
\noindent { (ii) $\langle C\rangle\ne 0\,$}:
\vskip 0.1cm

Here, the vacuum conditions read
\dis{
   &\epsilon(C+C^*)+\frac{\alpha}{f_a}M_P^2=0,\\
   &\frac{1}{1-\epsilon^2}\left|C+M_P^2\frac{W^*_{\bar C}}{W^*}\right|^2=3M_P^2 ,\\
   &\frac{1}{1-\epsilon^2}\left[\frac{W_{CC}}{W}M_P^2
   -\Big(\frac{W_C}{W}\Big)^2-\epsilon^2\right]=\frac{C^*+(W_C/W)}{C+ (W^*_{\bar C}/W^*)}   .
}
The left-hand side of the first equation is $G_A$, and  hence SUSY is unbroken in the $A$ direction. But, for $C$, we have
\dis{
  (C+C^*)=-\frac{\alpha}{\epsilon}\frac{M_P^2}{f_a}\simeq -\frac{\alpha}{\epsilon}10^{7}M_P\,,
  }
for $f_a\simeq 10^{-7}M_P$.
Hence, for the VEV of $C$ staying at the Planck scale, $\alpha$ needs to be small at O($10^{-7}$). Now, the fermion mass matrix is given by
  \dis{
  \frac{m}{m_{3/2}} &=\left(\begin{array}{cc}  1,  &\epsilon \\[1em]
      \epsilon, &
      \begin{array}{c}
      M_P^2\Big[\frac{W_{CC}}{W}-\Big(\frac{W_C}{W}\Big)^2 \Big]\\[0.6em]
      +\frac13\Big(\frac{C^*}{M_P}+M_P\frac{W_C}{W}\Big)^2
      \end{array}
\end{array}\right)\\
&=\left(\begin{array}{cc}  1  &\epsilon \\
      \epsilon & \epsilon^2\\
\end{array}\right)}
where the vacuum conditions are used to simplify the (22) element.

We should redefine fermions such that the kinetic terms are canonical, using Eq. (\ref{eq:Vexample}). The eigenvalues are again 0 and  $m_{3/2}$, confirming the correct goldstino mass eigenvalue. The coefficient of $m_{3/2}$ in the axino mass, $\xi_{\rm goldstino}$, is 1 again.

%%%%%%%%%%%%%%%%%%%%%%%%%%%%%%%%%%%%%%%%%%%%%%%%%%%%%%%%%%%%%%%%%%%%%%%%%%%%%%%%
\subsubsection{Case for $G_A\ne 0$ }

\vskip 0.3cm
In  {\it Case 1}, we considered
$K=\frac12(A+A^*)^2+CC^*+\epsilon(A+A^*)(C+C^*)$,
which gave $G^A \ne 0$ even though $G_A=0$ from the vacuum condition. To investigate the case of $G_A \ne 0$ from the beginning with no mixing with SUSY breaking coaxino $C$, Fig. \ref{fig:FAnonzero}, let us consider
 \dis{
 &K=f_a^2\Big[ e^{c_1(A+A^*)/f_a}+ e^{-c_2(A+A^*)/f_a} \Big]+CC^* \\
 &\quad W=e^{\alpha A/f_a}W(C).
 }
Then, with $A=0$, we obtain
\dis{
K_A&=f_a\Big[ c_1e^{c_1(A+A^*)/f_a}-c_2 e^{-c_2(A+A^*)/f_a} \Big]\\
&=f_a(c_1-c_2),\\
 K_C&=C^*
 }
and
 \dis{
 K_{AA}&=\Big[ c_1^2e^{c_1(A+A^*)/f_a}+ c_2^2e^{-c_2(A+A^*)/f_a} \Big]\\
 &=(c_1^2+c_2^2),
 \\
K_{AC}&=K_{CC}=0.
}
The  K\"ahler metric is given by
\dis{
K_{A{\bar A}}&=\Big[ c_1^2e^{c_1(A+A^*)/f_a}+ c_2^2e^{-c_2(A+A^*)/f_a} \Big]\\
&=(c_1^2+c_2^2)  \\
   K_{A\bar{C}}&=K_{C\bar{A}}=0,  \\
  K_{C\bar{C}}&=1,
}
and its inverse is given by
\dis{K^{A\bar{A}}=\frac{1}{K_{A\bar{A}}},~~K^{C{\bar C}}=1,~~ K^{A\bar{C}}=K^{C\bar{A}}=0.}
Since
\dis{
K_{AA\bar{A}}&=\frac{1}{f_a}\left[ c_1^3e^{c_1(A+A^*)/f_a}- c_2^3e^{-c_2(A+A^*)/f_a} \right]\\
&=\frac{1}{f_a}(c_1^3-c_2^3),
}
and other $K_{ij{\bar k}}\,$'s vanish, the only nonzero Christoffel symbol is
\dis{
\Gamma_{AA}^A=\frac{1}{f_a}\Big(\frac{c_1^3-c_2^3}{c_1^2+c_2^2}\Big).
}

Now, consider the SUSY breaking. The fermion mass matrix. $G_i$, the
barometer of SUSY breaking, is
\dis{&G_A=f_a(c_1-c_2)+\frac{\alpha}{f_a}M_P^2
\\
&G_C=C^*+\frac{W_C}{W}M_P^2.}

And, $G_{ij}$ in the mass matrix is given by
\dis{
&G_{AA}=(c_1^2+c_2^2)\\
&G_{AC}=0\\
&G_{CC}=M_P^2 \Big[ \frac{W_{CC}}{W}-\Big(\frac{W_C}{W}\Big)^2 \Big].
}

Note that $G_{AC}=0$. It is because of the form of the superpotential. If the superpotential is merely given by a sequestered form, $W=W^{(a)}(A)+W^{(c)}(C)$,
the contribution of the superpotential to $G_{AC}$ is
\dis{
\frac{W_{AC}}{W}-\frac{W_AW_C}{W^2}
=-\frac{W^{(a)}_AW^{(c)}_C}{W^2} \label{eq:73}
}
so that it does not vanish in general. However, the shift symmetry of $A$ in $G$ restricts the $A$ dependence of the superpotential to the form $  W(C)\, e^{\alpha A/\fa}$ so that $(W_{AC}/W)-(W_AW_C/W^2)$ vanishes. This seems not giving $m_{\tilde a}=2m_{3/2}$ of Ref. \cite{Nomura10}. The reason is the following.

The vacuum condition $G^{i{\bar j}}G_iG_{\bar j}=3M_P^2$ reads
\dis{
&\frac{1}{c_1^2+c_2^2}\Big(f_a(c_1-c_2)+\frac{\alpha}{f_a}M_P^2\Big)^2 \\ &+\Big|C^*+M_P^2\frac{W_C}{W} \Big|^2 =3M_P^2
}
and the conditions $G^j\nabla_iG_j+G_i=0 $ read

\dis{
&\frac{(c_1^3-c_2^3)}{(c_1^2+c_2^2)^2}\Big[(c_1-c_2)+\alpha\frac{M_P^2}{f_a^2}\Big]=2,
\\
&M_P^2\Big[\frac{W_{CC}}{W}-\Big(\frac{W_C}{W}\Big)^2\Big]
=-\frac{C^*+M_P^2W_C/W}{C+M_P^2W_{\bar C}^*/W^*}.
}

In fact, this is where the vanishing contribution of the superpotential to $G_{AC}$ comes in. Suppose $A=C=0$ and $K_i$ are negligible as assumed in \cite{Nomura10}. Then, $G_i \simeq M_P^2W_i/W$. Let $\nabla_iG_j=K_{ij}-\Gamma^k_{ij}G_k+M_P^2[(W_{ij}/W)-(W_iW_j/W^2)]\equiv X_{ij}-M_P^2(W_iW_j/W^2)$. For $i=A$ where we are interested in, the second vacuum condition reads
\dis{
0&=G^A\nabla_AG_A+G^C\nabla_AG_C+G_A\\
&=G^A[X_{AA}-M_P^2\frac{W_A^2}{W^2}]+G^C[X_{AC}-M_P^2\frac{W_AW_C}{W^2}]+G_A\\
&\simeq G^AX_{AA}+G^CX_{AC}-G_A[G^AG_A+G^CG_C]/M_P^2+G_A \label{eq:76}
}
where $G_i \simeq M_P^2W_i/W$ is used. Since $[G^AG_A+G^CG_C]=3M_P^2$ from the first vacuum condition, we have
\dis{
X_{AA} =2\frac{G_A}{G^A}-\frac{G^C}{G^A}X_{AC}.\label{eq:77}
}
The factor 2 in front of $(G_A/G^A)$ is the source of $2m_{3/2}$ of \cite{Nomura10}. For this factor, $W_AW_C/W^2$ in $\nabla_A G_C$ plays a crucial role but is vanishing with our potential, and we cannot use the first vacuum condition for the factor 2. So, the PQ invariance in the superpotential does not require $m_{\tilde{a}} \sim 2m_{3/2}$. In fact, Eq. (\ref{eq:76}) shows how factor 2 of Ref. \cite{Nomura10} comes about. The factor 2 in the RHS of (\ref{eq:77}) is its origin, $m=2m_{3/2}$. Note that it is based on the formalism of Eq. (\ref{eq:73}), the result of a sequestered superpotential. This does not hold in our case, as explained below Eq. (\ref{eq:73}). In our case, the superpotential of the form $e^{\alpha A/f_a}W(C)$ leads to  different relations on the mass term and other parameters from those of  Ref. \cite{Nomura10}, where the K\"ahler potential does not play a crucial role.

Even though our examples with one coaxino show $m_{\tilde a} \simeq m_{3/2}$, it does not imply that the axino mass should be the gravitino mass, as commented later below Eq. (\ref{eq:82}). Moreover, if we put mixing between $A$ and $C$ in the above K\"ahler potential, $(1/M_P^2)(A+A^*)CC^*$, the axino mass is not the exact gravitino mass. It may be larger or smaller than the gravitino mass, but no huge enhancement beyond of order 1: $m_{\tilde a}={\cal O}(1) \times m_{3/2}$.

Furthermore, vacuum conditions imply that
 \dis{f_a \le \frac{\sqrt3}{2}\frac{|c_1^3-c_2^3|}{(c_1^2+c_2^2)^{3/2}}M_P }
 which would be a criterion for reasonable PQ scale.

From the vacuum conditions, the fermion mass matrix is given by
\dis{\frac{m_{ij}}{m_{3/2}}&=\nabla_i G_j+\frac13 G_iG_j
\\
=&\left(\begin{array}{cc}  -(c_1^2+c_2^2)& 0  \\[0.5em]
      0 & -e^{2i\omega}
\end{array}\right)
+\frac13\left(\begin{array}{cc} G_A^2& G_AG_C \\[0.5em]
      G_AG_C & G_C^2
\end{array}\right)
}
where  $G_C \equiv {\cal C}e^{i\omega}$.
In the $G^i$ direction, we  confirm $m_{ij}G^j=0$, and we have the massless fermion in the $G^i$ direction, as expected.

The canonical kinetic terms can be obtained by rescaling,
\dis{V= \left(\begin{array}{cc}  \frac{1}{\sqrt{c_1^2+c_2^2}}& 0  \\
     0 &1
\end{array}\right).
}
Then, the physical mass matrix is given by
  \dis{\frac{m_{ab}}{m_{3/2}}=\left(\begin{array}{cc}  -1& 0  \\
     0 & -e^{2i\omega}
\end{array}\right)
+\frac{1}{3M_P^2} \left(\begin{array}{cc}  G_AG^A,&  G^AG_C  \\
     G^AG_C, &G_CG^C
\end{array}\right)}
since $G^A=(1/c_1^2+c_2^2)G_A$ and $G^C=G_C$. Then, the axino mass  Eq. (73) should be modified to
\dis{\frac{m_{\tilde a}}{m_{3/2}}&=-[1+e^{2i\omega}]+\frac{1}{3M_P^2}(G_AG^A+G_CG^C)
\\
&=-e^{2i\omega}\label{eq:82}
}
since $G_AG^A+G_CG^C=3M_P^2$. By the phase redefinition, we obtain $m_{\tilde a}=m_{3/2}$.

However, when $c_1=c_2 \equiv c$, the situation changes drastically. The vacuum condition requires $G_A=(\alpha/f_a)=0$ so that  axino does not break SUSY and is completely decoupled from the SUSY breaking field $Z=C$. So, the axino is still massless. It  may be a low energy effective description of $W=Z_1(S_1S_2-f_1^2)+Z_2(S_1S_2-f_2^2)$ discussed in \cite{Kim83}. Even though SUSY is broken in this case, the axion superfield does not appear in the superpotential. Since the axion superfield and the SUSY breaking superfield are decoupled in both K\"ahler potential (since canonical K\"ahler is assumed) and superpotential,  axino is massless since axion is massless.

Physics of the PQ symmetry breaking can introduce another scale of axino mass. Suppose we regard $\varphi$ as a dynamical superfiel such that  K\"ahler potential is given by, for example,
\dis{
K=\frac12 \varphi \varphi^*\Big[e^{c(A+A^*)/f_a}+e^{-c(A+A^*)/f_a}\Big].
}
One may assume that superpotential depends on $C$ and $\varphi$ separately, $W=e^{\alpha A/f_a}(W_1(\varphi)+W_2(C))$. In this case, we have to consider $3\times3$ fermion mass matrix in the basis $(\varphi, C, A)$ and find that three eigenvalues are zero, ${\cal O}(f_a)$, and ${\cal O}(m_{3/2})+{\cal O}(m_{3/2}^2/f_a)$ , respectively. The last case corresponds to the axino. When the leading order term ${\cal O}(m_{3/2})$ is suppressed from some accidental symmetry, the next leading order ${\cal O}(m_{3/2}^2/f_a)$ is the axino mass scale.

%%%%%%%%%%%%%%%%%%%%%%%%%%%%%%%%%%%%%%%%%%%%%%%%%%%%%%%%%%%%%%
\subsection{The KSVZ model}
In the KSVZ approach, one introduces the heavy quark fields $Q_L$ and $Q_R$ in the superpotential as  \cite{KSVZ79},
\dis{
W_{\rm KSVZ}=m^3\Theta\, e^{A/f_a} &+{f_Q} Q_L \overline{Q}_R \,
       \varphi  \, e^{A/f_a }  . \label{Wksvz}
}
The PQ symmetry is given near the $\epsilon$ point of Fig. \ref{fig:DMW}, with $\Gamma(\overline{Q}_L)=-1/2, \Gamma( {Q}_R)=-1/2,$ and $\Gamma(X)=1 $. Near $\epsilon$, there is no $\varphi$ type field. But near $\NDW$, $\Gamma(\overline{Q}_L)$ and $Q_R$ are not of the $\varphi$ type,  only $X$ is a $\varphi$ type field, and $Q$ obtains the heavy quark mass $m_Q=f_Q \langle\varphi(X)\rangle$.

It can be rephrased as follows.
The heavy quark interaction with $A$ after integrating out heavy scalars by $\varphi=\fa$, we have the interaction $m_QQ_L\overline{Q}_Re^{ A/f_a}$. Technically, we loose the PQ quantum number information of heavy quarks since they do not have a $\varphi$ type component but only the phase dependence by the original PQ charges. These phases can be rotated away by redefining the phases of $Q_L$ and $Q_R$. This heavy quark interaction with $A$ generates the two loop mass presented in Fig. \ref{fig:AxinoKSVZ} at the order of 10 GeV.

%\begin{widetext}
%%%%%%%%%%%%%%%%%%%%%%%%%%%%%%%%%%%%%%%%%%%%%%%%%%%%%%%%%%%%%%%%%%%%%%%
\begin{table}
\begin{center}
\begin{tabular}{|c|c|c|c|c|c|c|c|c|c| }
\hline
Model    & $S_1$ & $S_2$ & $Q_L$ & $\overline{Q}_R$
&$H_d$ &$H_u$& $q_L$& $D^c_R$ & $U^c_R$ \\
\hline
{\rm KSVZ}  & $1$&$-1$&$-\frac12 $&$-\frac12 $& 0& 0& 0&0&0\\
{\rm DFSZ}  & $1$&$-1$& 0 & 0&$-1 $ & $- 1$& $\ell$& $1-\ell$ &$1-\ell$\\
\hline
\end{tabular}
\caption{The PQ charge assignment $Q$. $Q_L$ and $\overline{Q}_R$ denote
new heavy  quark multiplets.   }
\label{table:PQcharge}
\end{center}
\end{table}
%%%%%%%%%%%%%%%%%%%%%%%%%%%%%%%%%%%%%%%%%%%%%%%%%%%%%%%%%%%%%%%%%%%%
%\end{widetext}
%%%%%%%%%%%%%%%%%%%%%%%%%%%%%%%%%%%%%%%%%%%%%%%%%%%%%%%%%%%%%%%%%%%%%%%%
\subsection{The DFSZ model}
In the DFSZ framework, the SU(2)$_L\times $U(1)$_Y$ Higgs doublets carry PQ charges
and thus the light quarks are also charged under \uonepq~ \cite{DFSZ81}.
The charge assignment is shown in Table~\ref{table:PQcharge}.
So, the superpotential is written as

\dis{
 W_{\rm DFSZ}=W_{\rm PQ}+ \frac{f_s}{\mplanck} S_1^2 H_d H_u,
}
where  $H_d H_u\equiv \epsilon_{\alpha\beta}H_d^\alpha H_u^\beta$. Integrating out $S_1$, we have
\dis{
 W_{\rm DFSZ}&= \mu  e^{2A/f_a}\varphi(H_d) \varphi(H_u)e^{-2A/f_a} \\
 &+f_u q_L e^{ {\ell}  \theta} u_R^c e^{(1-{\ell})\theta} \varphi(H_u)e^{-  A/f_a} \\
 &+f_d q_L e^{ {\ell} \theta} d_R^c e^{(1- {\ell} )\theta} \varphi(H_d)e^{-  A/f_a}   .\label{eq:DFZSsuper}
}
For the quarks, they do not contain the $\varphi$ type fields since they do not contribute to $V_a^2$ of Eq. (\ref{eq:Vasquare}) and their phase is just a phase parameter $\theta$. This $\theta$ can be removed by
redefining the phases of quarks, and we obtain
\dis{
 W_{\rm DFSZ}&= \mu   \frac{v_uv_d}{2}
 +( m_t t_L  t_R^c   + m_b b_L  b_R^c  +\cdots) e^{ A/f_a}  , \label{eq:DFZSsuperpotent}
}
with $\NDW=6$. Equation (\ref{eq:DFZSsuperpotent}) breaks the PQ symmetry and the axion coupling to stop is
\dis{
&M_P^4\ln \left(\frac{1}{2M_P^6}\mu v_uv_d m_t t_L t_R^c  e^{ A/f_a} +{\rm h.c.}
\right)\\
&\to \frac{1}{M_P^2}\mu v_uv_d m_t  |\langle t_L t_R^c \rangle| \cos\frac{a}{\fa}
}
whose coefficient is much smaller than the QCD instanton contribution $\Lambda_{\rm QCD}^4$. So we can neglect the constant term for the axion mass even though it breaks the PQ symmetry. Also, the constant term does not contribute to the axino mass. Including the SUSY breaking auxilliary field $\Theta$, we consider
\dis{
 W_{\rm DFSZ}&=\mu   \frac{v_uv_d}{2}+ \Big(m^3\Theta
 + m_t t_L  t_R^c   \\
  &+ m_b b_L  b_R^c+\cdots\Big)  e^{ A/f_a} .   \label{eq:DFZSsuperpotential}
}
Through the quark mass terms, we obtain the axino mass as shown in Fig. \ref{fig:AxinoKSVZ} with $Q$ replaced by the SM quarks. The SM quark mass is at most $m_t/m_Q\sim 10^{-9}$ smaller than that of the heavy quark and we obtain the axino mass in the range 10 eV. Only the SUSY breaking soft mass contribution $ m^3 \Theta$ can contribute to the axino mass.

 %%%%%%%%%%%%%%%%%%%%%%%%%%%%%%%%%%%%%%%%%
\section{Conclusion}

After properly defining the goldstino and axion multiplets, we presented the calculational scheme of the axino mass in the most general framework. For only two light superfields of goldstino and axino, we obtain $\xi_{\rm goldstino}$ of Eq. (\ref{eq:axinomassxis}). For $G_A=0$ where $G=K+\ln|W|^2$, we obtained
$m_{\tilde a}= m_{3/2}$ with the axino-gravitino mixing parameter  $\epsilon$  in the K\"ahler potential. For $G_A\ne 0$, we showed that the axino mass depends on the details of the K\"ahler potential.
But there is another parameter proportional to the gaugino masses, and we can take a wide range of the axino mass for cosmological applications \cite{Masiero84,RTW91,ChunKN92,CKR00,CKKR01,Steffen04, CKLS08,Huh09,ChoiCovi12,ChunLukas,NillesRaby82,Tamv82,Frere83,Strumia10,AxinoRevs}. If the gravity mediation is the dominant one, the axino mass is probably greater than the gravitino mass, but its decay to gravitino is negligible. Still, it softens the cosmological gravitino problem \cite{gravProb} somewhat as discussed in Ref. \cite{CKLS08}.

\medskip
%%%%%%%%%%%%%%%%%%%%%%%%%%%%%%%%%%%%%%%%%%%%%%%%%%%%%%%%%%%%%%%%%%%%%%%%%%%%%%%%%%%
\acknowledgments{This work is supported in
  part by the National Research Foundation (NRF) grant funded by the
  Korean Government (MEST) (No. 2005-0093841). }

%%%%%%%%%%%%%%%%%%%%%%%%%%%%%%%%%%%%%%%%%%%%%%%%%%%%%%%%%%%%%%%%%%%%%%%%%%%%%%%%

%%%%%%%%%%%%%%%%%%%%%%%%%%%%%%%%%%%%%%%%%%%%%%%%%%%%%%%%%%%%%%%%%%%%%%%

\def\prp#1#2#3{Phys.\ Rep.\ {\bf #1} (#3) #2}
\def\rmp#1#2#3{Rev. Mod. Phys.\ {\bf #1} (#3) #2}
\def\anrnp#1#2#3{Annu. Rev. Nucl.
Part. Sci.\ {\bf #1} (#3) #2}
\def\npb#1#2#3{Nucl.\ Phys.\ {\bf B#1} (#3) #2}
\def\plb#1#2#3{Phys.\ Lett.\ {\bf B#1} (#3) #2}
\def\prd#1#2#3{Phys.\ Rev.\ {\bf D#1} (#3) #2}
\def\prl#1#2#3{Phys.\ Rev.\ Lett.\ {\bf #1} (#3) #2}
\def\jhep#1#2#3{J. High Energy Phys.\ {\bf #1} (#3) #2}
\def\jcap#1#2#3{J. Cosm. and Astropart. Phys.\ {\bf #1} (#3) #2}
\def\zp#1#2#3{Z.\ Phys.\ {\bf #1} (#3) #2}
\def\epjc#1#2#3{Euro. Phys. J.\ {\bf #1} (#3) #2}
\def\ijmp#1#2#3{Int.\ J.\ Mod.\ Phys.\ {\bf #1} (#3) #2}
\def\mpl#1#2#3{Mod.\ Phys.\ Lett.\ {\bf #1} (#3) #2}
\def\apj#1#2#3{Astrophys.\ J.\ {\bf #1} (#3) #2}
\def\nat#1#2#3{Nature\ {\bf #1} (#3) #2}
\def\sjnp#1#2#3{Sov.\ J.\ Nucl.\ Phys.\ {\bf #1} (#3) #2}
\def\apj#1#2#3{Astrophys.\ J.\ {\bf #1} (#3) #2}
\def\ijmp#1#2#3{Int.\ J.\ Mod.\ Phys.\ {\bf #1} (#3) #2}
\def\apph#1#2#3{Astropart.\ Phys.\ {\bf B#1}, #2 (#3)}
\def\mnras#1#2#3{Mon.\ Not.\ R.\ Astron.\ Soc.\ {\bf #1} (#3) #2}
\def\apjs#1#2#3{Astrophys.\ J.\ Supp.\ {\bf #1} (#3) #2}
\def\aipcp#1#2#3{AIP Conf.\ Proc.\ {\bf #1} (#3) #2}

%%%%%%%%%%%%%%%%%%%%%%%%%%%%%%%%%%%%%%%%%%%%%%%%%%%%%%%%%%%%%%%%%%%%%%%


\begin{thebibliography}{99}



%%%%%%%%%%%%%%%%%%%%%%%%%%%%%%%%%%%%%%%%%%%%%%%%%%%%%%%%%%%%%%%%%%

\bibitem{CKR00}  L.~Covi, J.~E.~Kim and L.~Roszkowski,
%  {\it Axinos as CDM},
  \prl{82}{4180}{1999}  [hep-ph/9905212].

\bibitem{CKKR01}
  L.~Covi, H.~B.~Kim, J.~E.~Kim and L.~Roszkowski,
%  {\it Axinos as dark matter},
  \jhep{0105}{033}{2001}  [hep-ph/0101009].

\bibitem{CKLS08}  K.-Y. Choi, J. E. Kim, H. M. Lee, and O. Seto, \prd{77}{123501}{2008} % Neutralino dark matter from heavy axino decay
    [arXiv: 0801.0491[hep-ph]].

\bibitem{Steffen04}  A. Brandenburg and F. D. Steffen, \jcap{0408}{008}{2004}  [hep-ph/0405158].

\bibitem{ChunKN92} E. J. Chun, J. E. Kim and H. P. Nilles, \plb{287}{123}{1992}.

\bibitem{Huh09} J.-H. Huh and J. E. Kim, \prd{80}{075012}{2009}
[arXiv:0908.0152 [hep-ph]].

\bibitem{ChoiCovi12}   K.-Y. Choi, J. E. Kim, L. Covi, and L. Roszkowski, \jhep{}{}{2012}
    [arXiv:1108.2282 [hep-ph]].

\bibitem{Masiero84} J. E. Kim, A. Masiero and D. V. Nanopoulos, \plb{139}{346}{1984}.

\bibitem{RTW91}
K.~Rajagopal, M.S.~Turner and F.~Wilczek,
%{\it Cosmological implications of axinos},
\npb{358}{1991}{447}.

\bibitem{ChunLukas} E. J. Chun and A. Lukas, \plb{357}{43}{1995} [arXiv:hep-ph/9503233].

\bibitem{NillesRaby82} H. P. Nilles and S. Raby, \npb{198}{102}{1982}.

\bibitem{Tamv82} K. Tamvakis and D. Wyler, \plb{112}{451}{1982}.

\bibitem{Frere83} J. M. Frere and J. M. Gerard, Lett. Nuovo Cim. {\bf 37}, 135 (1983).

\bibitem{Strumia10} A. Strumia, \jhep{1006}{036}{2010} [arXiv:1003.5847 [hep-ph]].

\bibitem{AxinoRevs} For a recent review, see,  L. Covi and J. E. Kim, New J. of Phys. {\bf 11} (2009) 105003
    [arXiv: 0902.0769[astro-ph/CO]]; H. Baer and A. D. Box, \epjc{C68}{523}{2010} [arXiv:0910.0333 [hep-ph]].

\bibitem{Kim83} J. E. Kim, \plb{136}{78}{1984}.

\bibitem{KimRMP10} For a recent review, see, J. E. Kim and G. Carosi, \rmp{82}{557}{2010} [arXiv: 0807.3125[hep-ph]].

\bibitem{KimPRP87} J. E. Kim, \prp{150}{1}{1987}; K. Choi and J. E. Kim, \prd{32}{1828}{1985} and \prl{55}{2637}{1985}.

\bibitem{GiuMas88} G. G. Giudice and A. Masiero, \plb{206}{480}{1988}.

\bibitem{KimNilles} J. E. Kim and H. P. Nilles,  \plb{138}{150}{1984}.

\bibitem{Deser77} S. Deser and B. Zumino, \prl{38}{1433}{1977}.

\bibitem{Bae11} K. J. Bae, K. Choi, and S. H. Im, \jhep{1108}{065}{2011}   [arXiv:1106.2452].

\bibitem{KimNillSeo12} J. E. Kim, H. P. Nilles and M.-S. Seo, arXiv:1201.6547.

\bibitem{Yamaguchi11} K. S. Jeong, Y. Shoji, and M. Yamaguchi, arXiv: 1112.1014.

\bibitem{KSVZ79} J. E. Kim, \prl{43}{103}{1979}; M. A. Shifman, V. I. Vainstein, V. I. Zakharov, \npb{166}{4933}{1980}.

\bibitem{DFSZ81} M. Dine, W. Fischler and M. Srednicki, \plb{104}{199}{1981};  A. P. Zhitnitskii, Sov. J. Nucl. Phys. {\bf 31}, 260 (1980).

\bibitem{Svrcek06} P. Svrcek and E. Witten, \jhep{0606}{051}{2006}  [arXiv:hep-th/0605206].

\bibitem{ChoiK85} K. Choi and J. E. Kim, \plb{154}{393}{1985}.

\bibitem{AnomMed02} L. Randall and R. Sundrum, \npb{557}{79}{1999}  [arXiv:hep-th/9810155] J. Bagger, \jhep{0004}{009}{2000}  [arXiv:hep-th/9911029].

%\bibitem{AnomMed02} L. Randall and R. Sundrum, \npb{557}{79}{1999}  [arXiv:hep-th/9810155].%; J. Bagger, \jhep{0004}{009}{2000}  [arXiv:hep-th/9911029].

\bibitem{Yamaguchi02} N. Abe, T. Moroi and M. Yamaguchi, \jhep{0201}{010}{2002} [arXiv:hep-ph/0111155].

\bibitem{Polonyi77} J. Polonyi, Budapest preprint KFKI-1977-93 (1977).

\bibitem{Nomura10}  C. Cheung, Y. Nomura  and J. Thaler, \jhep{03}{073}{2010}  [arXiv:1002.1967].

\bibitem{gravProb} J. Ellis, J. E. Kim and D. V. Nanopoulos, \plb{145}{181}{1984}; M. Kawasaki, K. Kohri and T. Moroi, \prd{71}{083502}{2005} [astro-ph/0408426].

%\cite{Higaki:2011bz}
%\bibitem{Higaki:2011bz}
%  T.~Higaki and R.~Kitano,
  %``On Supersymmetric Effective Theories of Axion,''
%  arXiv:1104.0170 [hep-ph].
  %%CITATION = ARXIV:1104.0170;%%


%\bibitem{WessBag92} J. Wess and J. Bagger, {\it Supersymmetry and
%    Supergravity} (Princeton Univ. Press, 1992), Eq.~(6.11).

\end{thebibliography}
\end{document}